\documentclass[preprint]{aastex}
\usepackage{apjfonts,amsmath,natbib,graphicx,euscript,epsf,apjfonts,color,ulem}
\newcommand{\vect}[1]{\mathbf{#1}}

\newcommand{\rhoc}{\rho_{\rm c}}

\newcommand{\ltsim}{\protect\raisebox{-0.5ex}{$\:\stackrel{\textstyle <}{\sim}\:$}}
\newcommand{\gtsim}{\protect\raisebox{-0.5ex}{$\:\stackrel{\textstyle >}{\sim}\:$}}

\shorttitle{RMHD simulation of Star Formation}
\shortauthors{Tomida et al.}
\begin{document}
\title{Radiation Magnetohydrodynamics Simulation of Proto-Stellar Collapse: Two-Component Molecular Outflow}

\author{Kengo Tomida\altaffilmark{1,2}, Kohji Tomisaka\altaffilmark{1,2}, Tomoaki Matsumoto\altaffilmark{3}, Ken Ohsuga\altaffilmark{1,2}, Masahiro N. Machida\altaffilmark{2}, and Kazuya Saigo\altaffilmark{2}}

\altaffiltext{1}{Department of Astronomical Science, The Graduate University for Advanced Studies (SOKENDAI), Osawa, Mitaka, Tokyo 181-8588, Japan; tomida@th.nao.ac.jp, tomisaka@th.nao.ac.jp, ken.ohsuga@nao.ac.jp}
\altaffiltext{2}{National Astronomical Observatory of Japan, Osawa, Mitaka, Tokyo 181-8588, Japan; masahiro.machida@nao.ac.jp, saigo@th.nao.ac.jp}
\altaffiltext{3}{Faculty of Humanity and Environment, Hosei University, Fujimi, Chiyodaku, Tokyo 102-8160, Japan; matsu@hosei.ac.jp}

\begin{abstract}
We perform a three-dimensional nested-grid radiation magneto-hydrodynamics (RMHD) simulation with self-gravity to study the early phase of the low-mass star formation process from a rotating molecular cloud core to a first adiabatic core just before the second collapse begins. Radiation transfer is handled with the flux-limited diffusion approximation, operator-splitting and implicit time-integrator. In the RMHD simulation, the outer region of the first core attains a higher entropy and the size of first core is larger than that in the magnetohydrodynamics simulations with the barotropic approximation. Bipolar molecular outflow consisting of two components is driven by magnetic Lorentz force via different mechanisms, and shock heating by the outflow is observed. Using the RMHD simulation we can  predict and interpret the observed properties of star-forming clouds, first cores and outflows with millimeter/submillimeter radio interferometers, especially the Atacama Large Millimeter/submillimeter Array (ALMA).
\end{abstract}

\section{Introduction}
Radiation transfer plays a critical role in star formation and affects the structure of accretion flow and the resulting adiabatic cores even in a low-mass regime. However multi-dimensional radiation hydrodynamics (RHD) simulations have been rarely performed due to their high computational cost. Therefore, the barotropic approximation, which omits radiation transfer and simplifies the thermal evolution of the gas, is widely used in multi-dimensional simulations. However, recent advancement in the computer technology and development of numerical techniques enable us to incorporate radiation transfer into a multi-dimensional magnetohydrodynamics~(MHD) simulation within reasonable computational time, using a moment method with simplified closure relations. We performed RMHD simulations of proto-stellar collapse with our newly developed numerical code and clarified the difference between RMHD and MHD with the barotropic approximation.

RHD and RMHD simulation codes are being actively developed and some of them are used to study star formation processes. For example, \citet{yk95} solved radiation hydrodynamics by using the flux limited diffusion (FLD) approximation on 2D nested-grids. \citet{wb06} performed smoothed particle hydrodynamics simulations with FLD radiation transfer and showed different thermal evolution between models with and without radiation transfer. Orion AMR (Adaptive Mesh Refinement) RHD code developed by \citet{orion} is used to study the main accretion phase of low-mass star formation \citep{off09} and high-mass star formation \citep{krm09}. Recently AMR RMHD simulations based on RAMSES were reported by \citet{com09}, who studied the influence of radiation on fragmentation.

 Magnetic fields and self-gravity are also the key physical processes in star formation because they assume leading roles in the transfer of angular momentum which dominates the global evolution of a cloud. Magnetic braking and outflow driven by magnetic fields efficiently carry the initial angular momentum away from the cloud \citep{tmsk02,mim06}. If the cloud rotates very fast, it will fragment through self-gravitational instability \citep{mtmi08}. Since the structure and fate of the core strongly depend on its rotation and the efficiency of angular momentum transport, 3D MHD simulations with self-gravity are required.

 Molecular outflows from young stellar objects are universally observed with millimeter / submillimeter radio instruments (see \citealt{pp5} and references therein). In those observations some of outflows show complicated structure which can be interpreted as consisting of multiple components \citep{sg09}. They are strong evidences that magnetic fields play a crucial role in angular momentum transport during protostellar collapse. The argument about the driving mechanism of molecular outflow is not yet settled. We propose that the molecular outflow is driven via magnetic fields in the first core \citep{mim06}; however, some groups believe that the outflow is entrained by a high velocity jet from the central protostar \citep{vel07}. Next generation radio interferometric telescopes like the Atacama Large Millimeter/submillimeter Array (ALMA) have the potential to reveal the on-going star formation sites directly and contribute to our understanding about overall star formation processes and such important physical mechanisms.

In this letter, we report the results of a low-mass star formation simulation obtained with our newly developed 3D nested-grid self-gravitational RMHD code. We performed very deep nested-grid simulation of proto-stellar collapse just before the second collapse. In our simulation, a two-component outflow is driven from the first core. 
 In this letter, we concentrate on disscussion about these core and outflow including comparison between the RMHD simulation and the barotropic approximation while another paper is under preparation. The structure of this letter is as follows: In \S 2 we describe our numerical method and model setup. We show numerical results in \S 3, and \S 4 is devoted to discussions and conclusions.

\section{Method and Model}

We solve 3D self-gravitational RMHD equations on nested-grids. Equations and quantities for the radiation are defined in the comoving frame of the fluid. We treat only the zeroth order moment equations of radiation transfer using the gray (i.e., equations are integrated over frequency assuming a blackbody spectrum of a local temperature) FLD approximation proposed by \citet{lp81} with \citet{lev84} closure relations. We assume ideal MHD with no resistivity. The equation of state (EOS) of the gas is also assumed to be ideal, where the gas mainly consists of hydrogen molecules and the adiabatic index $\gamma$ is set to $7/5$ throughout the simulation, which is better to trace the realistic thermal evolution than $\gamma=5/3$ \citep{stm09}. The gas and the dust are assumed to have the common local temperature.

 Here we do not describe the implementation of our code in detail but mention some points about the treatment of radiation transfer. For the MHD and self-gravity parts, see \citet{mcd05a,mcd05b}. Our method for updating radiation and gas energy is similar to ZEUS-MP code \citep{zeusmp}, although our operator-splitting and treatment of the MHD and radiation parts are different from theirs. While our MHD and self-gravitaty parts are advanced explicitly with a second-order scheme, the radiation part is advanced implicitly with the first-order backward Euler method for stable time-integration. Newton-Raphson iteration is used to solve the resulting non-linear equations. The sparse matrix appearing here is very large and not diagonally dominant because of the strong non-linearity of the system; hence, a fast and robust iterative solver is required. We use a stabilized biconjugate gradient (BiCGStab) solver with a multi-color modified incomplete LU decomposition (ILU(0)) preconditioner, which is stable and efficient.

We determine the timestep by the Courant-Friedrich-Levy (CFL) criterion derived for the MHD part. All the grids have this common timestep and are advanced synchronously although individual (or asynchronous) timesteps are generally used in AMR simulations. In usual MHD simulations using individual timesteps, boundary values are constructed by time-interpolation of the values in the coarser level. However, if we use an implicit time-integrator and far larger timestep than that determined by the CFL condition for radiation transfer (i.e., $\Delta t \gg \Delta t_{RT}=\Delta x/c$), then this time-interpolation is not adequate, at least in principle.

We use a compiled table of opacity adopted from \citet{fer05} and \citet{semenov}. We use \citet{semenov} for the Rosseland mean opacity, and we smoothly combine the two for the Planck mean. That is, we use \citet{fer05} in high temperature ($T \gtsim 1000 {\rm K}$) and \citet{semenov} in low temperature because the Planck mean opacity of \citet{semenov} is quite lower than other opacity tables in high temperature \citep{fer05}. We adopt the surface formula with the limiter proposed by \citet{hg03} to evaluate radiation energy flux at the cell interface.

We take a Bonnor-Ebert sphere of 10K with central density $\rhoc=1.0\times 10^{-19}{\rm g\,cm^{-3}}$ for the initial condition. Initial rotation and magnetic fields are given uniformly along {\it z}-axis, $\omega=0.1/t_{ff}\simeq 1.5\times 10^{-14} \  {\rm sec^{-1}}$ and $B_z=1.1 {\rm \mu G}$. The outermost boundary values are fixed to keep their initial values and no geometric symmetry is assumed. The number of grid points in each level of nested-grids is $64^3$. The size of the finer grid is half of the coarser grid, and the finer grid is placed around the center of the simulation box self-similarly. The simulation starts with 5 levels of nested-grids, and finer grids are generated adaptively to resolve the local Jeans length with 32 meshes not to induce artificial fragmentation \citep{trlv97}. This is also because insufficient resolution causes the gas entropy to be overestimated near the center of the core. At the end of the simulation 18 levels of nested-grids are generated and the finest resolution is $\Delta x \sim 0.009 {\rm AU}$. We stop our simulation when the gas temperature of the central region reaches 2000K at which temperature hydrogen molecule starts to dissociate and the second collapse begins.

\section{Results}
\subsection{Overview}

\begin{figure}[p]
\scalebox{0.5}{\includegraphics{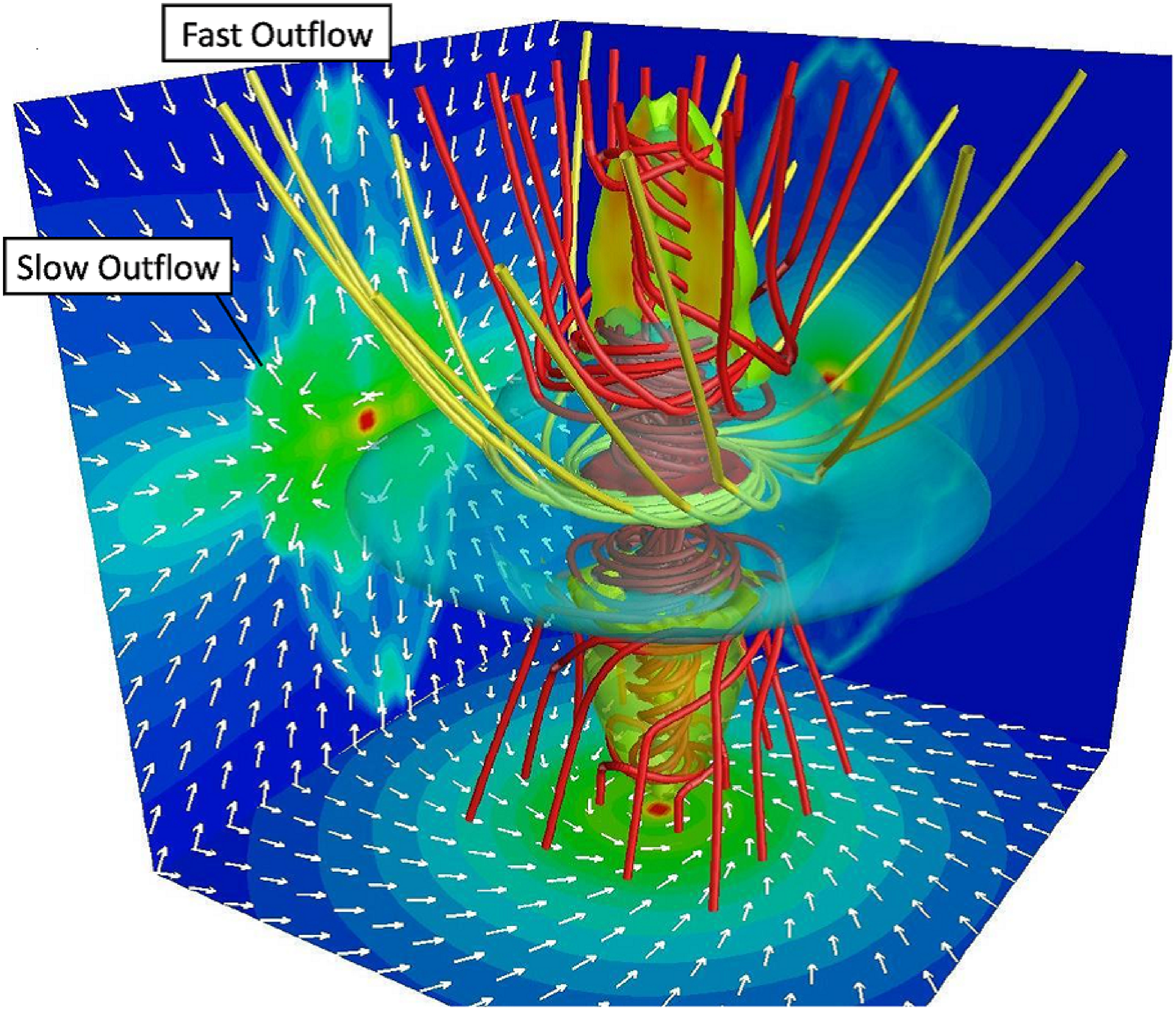}}
\caption{3D structure of the first core and outflow in level $L=10$. The left and bottom panels are density profiles and the right panel shows the temperature distribution. The cyan surface is a density isosurface. The fast outflowing region ($v_z > 0.3 \ {\rm km \ s^{-1}}$) is also visualized with volume rendering. Red and yellow lines are the magnetic field lines.}
\label{f1}
\end{figure}

Figure~\ref{f1} is a typical 3D bird's-eye view of the first adiabatic core and the outflow $\sim 500 \  {\rm yr}$ after the first core formation. There coexist two components of outflow: well collimated fast outflow (associated with red magnetic field lines in Figure~\ref{f1}) and slow outflow with a large opening angle (associated with yellow magnetic field lines). The former is driven by magnetic pressure and the latter by magneto-centrifugal force \citep{bp82}. \citet{tmsk02} shows that the magnetic pressure mode typically appears in case of weak magnetic fields, while the magneto-centrifugal mechanism appears in case of relatively strong magnetic fields.

\begin{figure}[p]
\scalebox{0.4}{\includegraphics{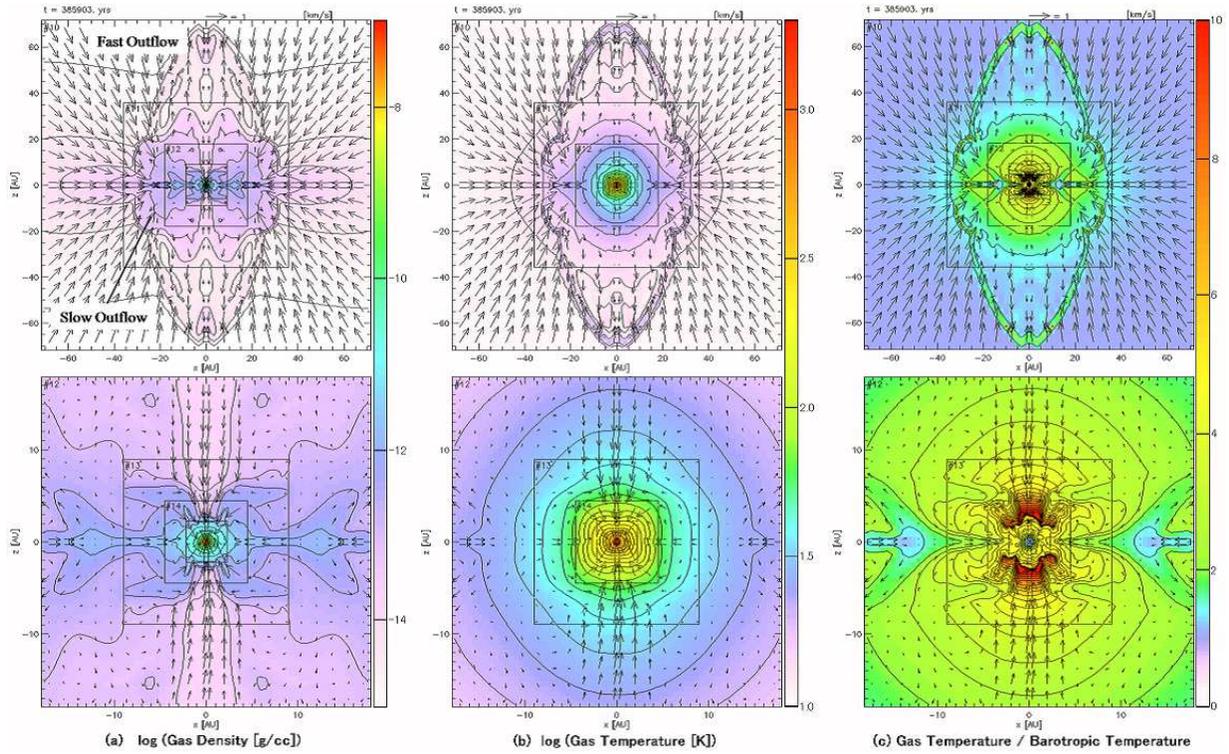}}
\caption{Vertical slices of (a) gas density, (b) gas temperature, and (c) gas temperature to the barotropic temperature ratio. The upper row is level $L=10$ corresponding to a scale of $\sim 150 {\rm AU}$ and the lower row is level $L=12$ corresponding to a scale of $\sim 40 {\rm AU}$. The projected velocity of the gas is overplotted with arrows.}
\label{f2}
\end{figure}

 Vertical slices of the outflow scale (level $L=10$; upper row) and core scale (level $L=12$; lower row) are shown in Figure~\ref{f2}. Hereafter we discuss only vertical slices since the process occurs nearly axisymmetric in this case. Although the density distribution shows the complicated structure as a result of MHD processes, the temperature distribution is almost spherically symmetric. The gas is heated up by the radiation before it enters the shock of the first core. This means that the temperature distribution in the outer region is dominated by the radiation from the central hot region. Shock heating up to $T\sim 30 {\rm K}$ at the edge of outflow is observed. This picture is considerably different from that of the barotropic approximation used in previous studies, in which the temperature is determined only by the local gas density.

 The difference between the RMHD simulation and the barotropic approximation is clearly seen in the right column (c) in Figure~\ref{f2} where the ratio of the gas temperature obtained in the RMHD simulation to the barotropic temperature given by the local gas density is plotted. In the RMHD simulation, the gas tends to attain the temperature typically 2-3 times higher than that in barotropic approximation around $\sim 10 {\rm AU}$ from the center, although the two models are very close in the innermost region of the core ($\ltsim 0.1 {\rm AU}$). The difference is most striking just above the shock at the surface of the first core because of pre-shock heating by radiation from the core. Here we use the following polytropic relation to evaluate the barotropic temperature:
\begin{equation}
T = 10 \left[\frac{\max(\rho,\rhoc)}{\rhoc}\right]^{\gamma-1}\hspace{1em} {\rm K ,}
\end{equation}
where $\rhoc=2.0\times 10^{-13}{\rm g\,cm^{-3}}$ is the critical density and $\gamma=7/5$ is the adiabatic index of diatomic molecular gas. The parameters used here are chosen to trace the thermal evolution track of the central region in a spherical RHD simulation by \citet{mi00}.

\subsection{Two-Component Outflow}
\begin{figure}[p]
\scalebox{0.66}{\includegraphics{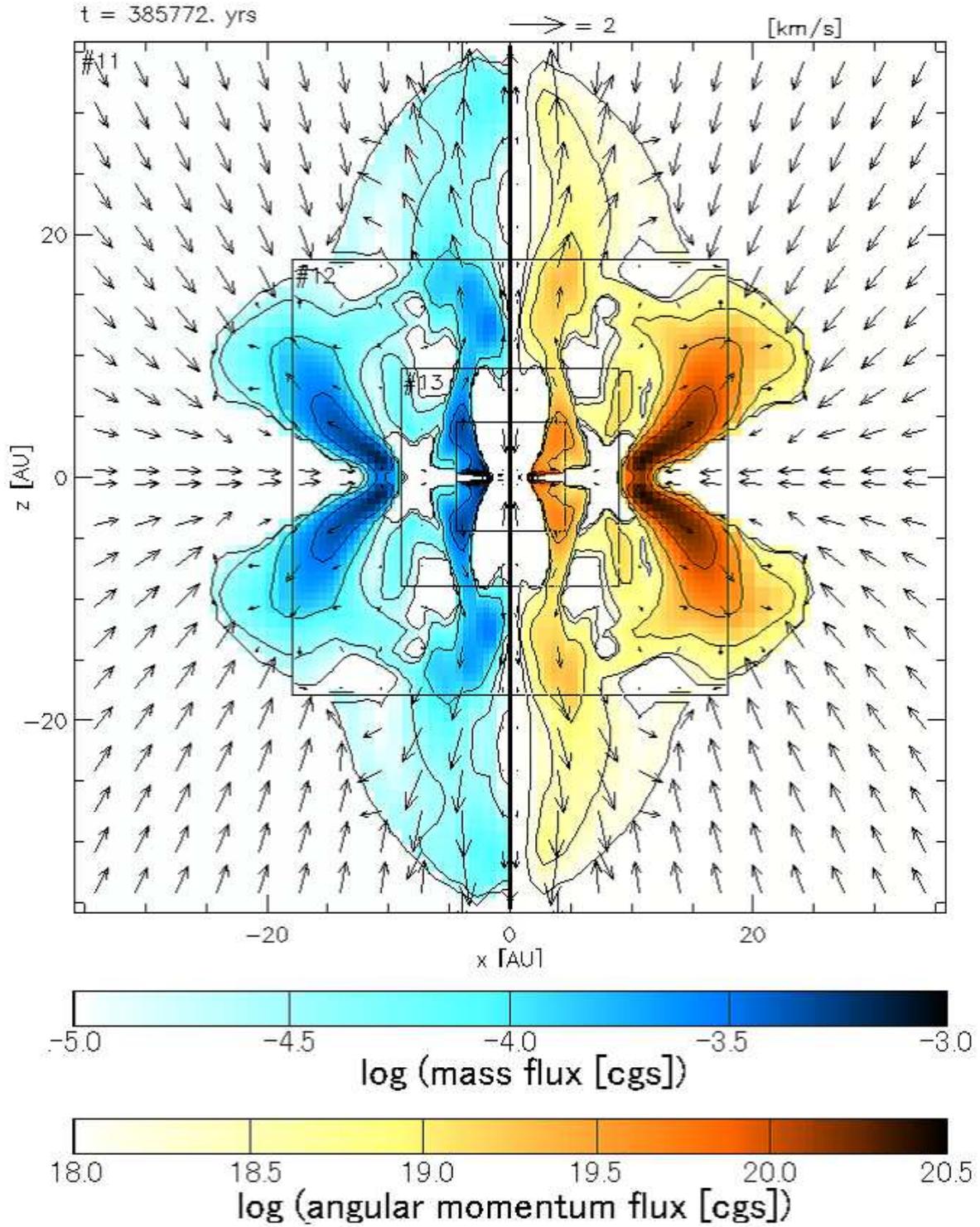}}
\caption{Outward mass flux (left) and outward angular momentum flux (right) in level $L=11$ at the early driving phase of the outflow are graphed. Two components of the outflow are clearly observable.}
\label{f3}
\end{figure}

We visualize the outgoing mass flux and angular momentum flux defined as $|\rho \vect{v_p}|$ and $|\rho \vect{v_p} |\vect{r}\times\vect{v_t}||$ where $v_r > 0$ respectively where $\vect{v_p} (\vect{v_t})$ is poloidal (toroidal) velocity at the early driving phase ($\sim 350 \  {\rm yr}$ after the first core formation) of outflow in Figure~\ref{f3}. In the figure we can recognize two components of outflow driven by different mechanisms: the outer magneto-centrifugal mode and the inner magnetic pressure mode.

 The inner magnetic pressure mode has a relatively high velocity $\sim 2 \ {\rm km \  s^{-1}}$ and is well collimated. The front of this outflow gets hot through shock heating due to its high Mach number. However, the outer, slow outflow has the mass and angular momentum flux, one order of magnitude larger, since the gas in the inner region has a relatively small angular momentum. This trend is enhanced as the outflow evolves. Accordingly we can say that the outer magneto-centrifugal mode is more important for the kinematic structure of the accretion flow as a carrier of angular momentum; however, the fast, inner outflow will stand out more for observation because it travels a long distance and achieves a high temperature.

\subsection{Structure of the First Core}

\begin{figure}[p]
\scalebox{1.35}{\includegraphics{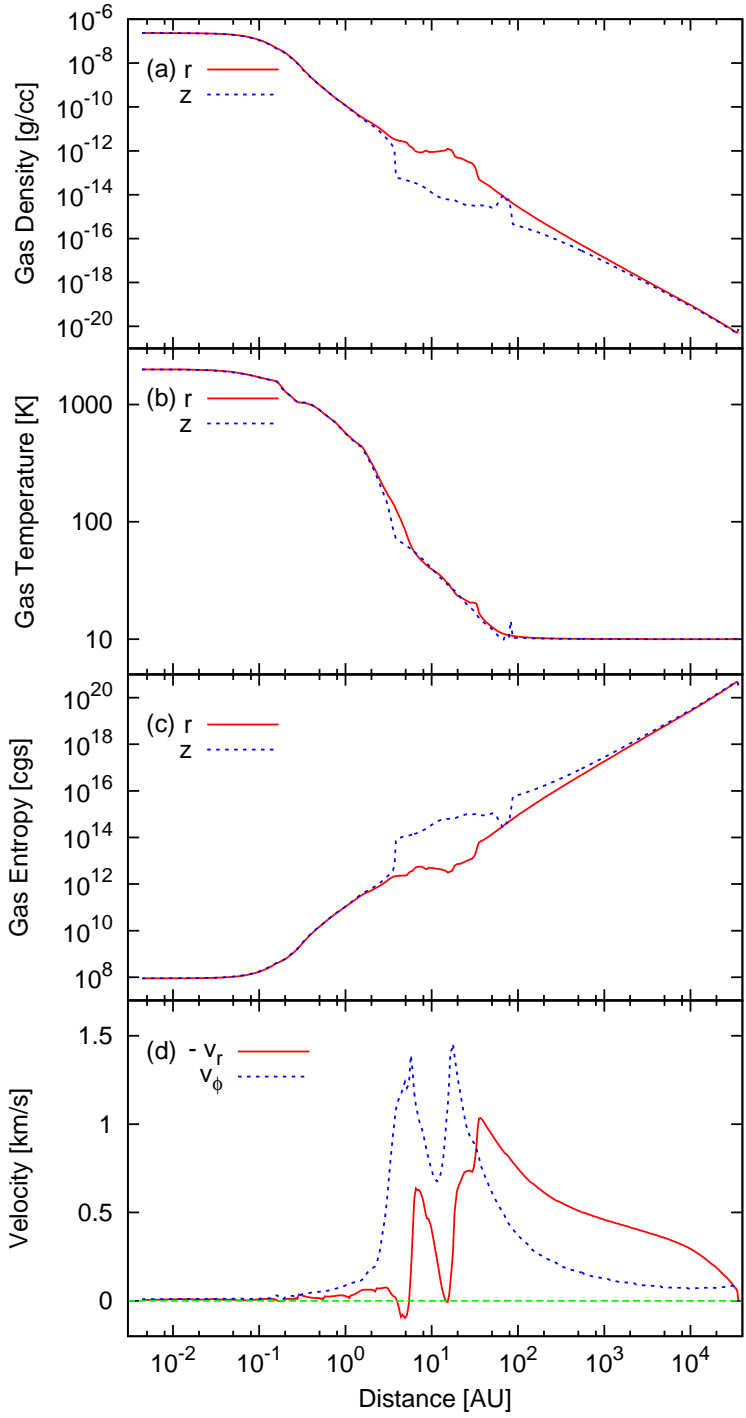}}
\caption{Plots of (a) gas density, (b) gas temperature, and (c) gas entropy ($K=P/\rho^\gamma$) in the disk midplane ({\it r}; red solid line) and along the rotational axis ({\it z}; blue dashed line), and (d) infalling/rotational ($-v_r$; red solid line / $v_{\phi}$; blue dashed line) velocities in the disk midplane at the end of the simulation.}
\label{f4}
\end{figure}

We can see a more detailed structure of the collapsing cloud from Figure~\ref{f4} where we plot the distribution of physical quantities at the end of the simulation, just before the second collapse starts  ($\sim 650 \  {\rm yr}$ after the first core formation). Panels (a)-(c) show the distribution of physical quantities in the disk midplane and along the rotational axis. At this time the fast outflow reaches $z \sim 85 \  {\rm AU}$. The surface of the core is located at $r \sim 35 {\rm AU}$ in the disk midplane and $z \sim 4 {\rm AU}$ along the rotational axis. This height of the first core is larger than that with the barotropic approximation, which is typically $\sim 1 {\rm AU}$ \citep{saigo08}, because the outer region of the first core achieves a higher entropy in RMHD simulation than in the barotropic approximation. The core with a higher entropy can support more mass even with the same central density; therefore, the first core lives longer in RMHD under the same accretion rate, which is mostly determined by the initial condition. Along the {\it z}-axis there is a high entropy radiative precursor outside the core ($z \gtsim 5 {\rm AU}$) seen in (b) and (c). The central region within $\sim 2 {\rm AU}$ from the center seems nearly spherically symmetric due to the slow rotation because of efficient magnetic braking.

The structure of the disk and the behavior of angular momentum transport is visible in panel (d) of Figure~\ref{f4} and Figure~\ref{f2}. The profiles of infalling velocity $-v_r$ and rotational velocity $v_{\phi}$ are plotted in panel (d) of Figure~\ref{f4}. The infalling gas initially decelerates at the weak shock near $\sim 35 {\rm AU}$ associated with the first core. There exist two centrifugal barriers, and the outer one near $\sim 20 {\rm AU}$ corresponds to the driving region of the slow and wide outflow driven via the magneto-centrifugal force. This outflow carries off angular momentum of the gas efficiently and the gas falls radially. Nonetheless in this case the initial magnetic fields are not so strong, the gas spins up again as it falls and hits the inner centrifugal barrier near $\sim 5 {\rm AU}$. Then, the magnetic field lines are wound up tightly (also due to weak initial magnetic fields) and magnetic pressure launches the fast, well collimated outflow.

\begin{figure}[p]
\scalebox{2}{\includegraphics{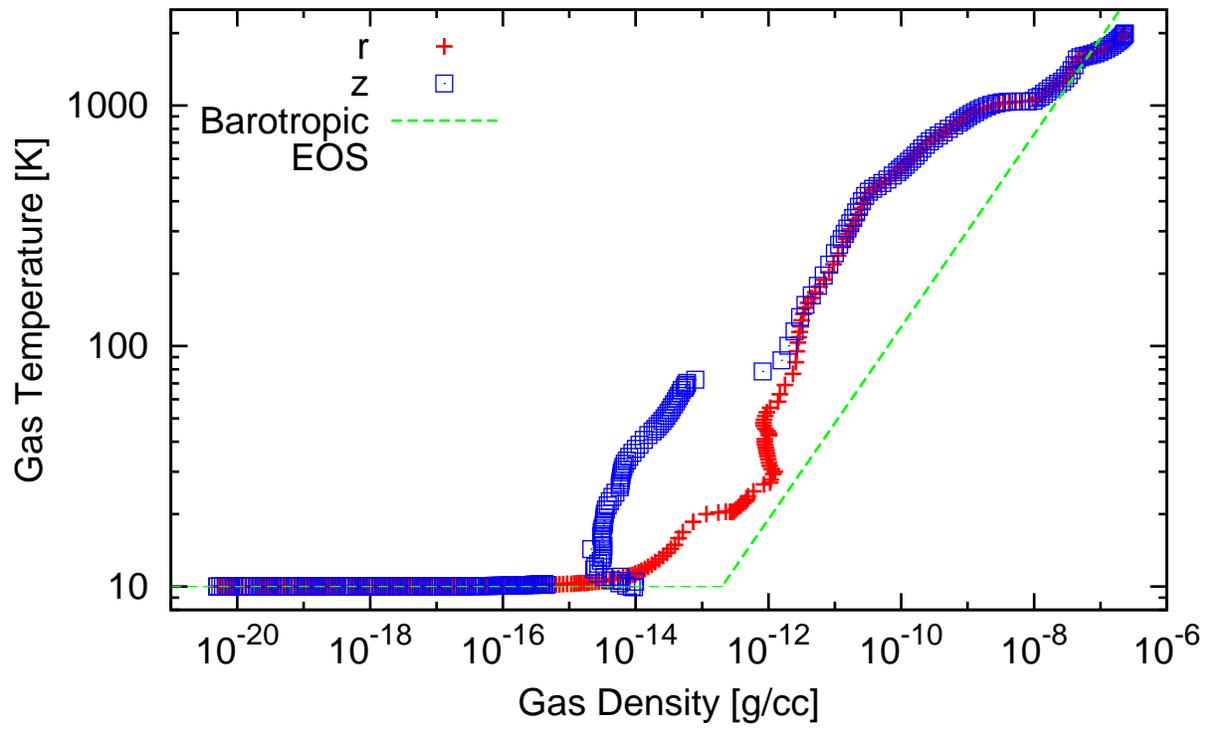}}
\caption{Distribution of thermal properties of the gas at just before the second collapse starts ($\sim 650 \  {\rm yr}$ after the first core formation) in the $\rho-T$ plane in the disk midplane ({\it r}; red crosses) and along the rotational axis ({\it z}; blue squares). The green dashed line is the barotropic EOS used in previous simulations.}
\label{f5}
\end{figure}

The variety of the gas's thermal behavior is shown in the $\rho - T$ plane in Figure~\ref{f5}. The distribution in this plane indicates a strong dependence on the geometry. We can see again that the gas tend to be hotter in the RMHD simulation than in the barotropic approximation. Although this was already pointed out by \citet{wb06}, our simulation can resolve more detailed features. The jump around $\rho\sim 10^{-13} {\rm g \, cm^{-3}}$ is caused by the shock at the surface of the first core. This density is very close to the critical density in the barotropic approximation. This shock is nearly isothermal since radiation can transfer thermal energy produced at the shock to the upstream efficiently. Radiation transfer is not so efficient beyond this critical density and this forms the first adiabatic core. When the temperature reaches $\sim 1500 {\rm K}$ the evolution track in the $\rho - T$ plane gets shallow. This is because all the components of the dust evaporate and the opacity drops sharply at this temperature, and then radiation transfer becomes effective again. The structure of the first core is slightly modified by this effect but not drastically since the gas soon becomes optically thick and adiabatic again. We emphasize that radiation transfer is required to handle the realistic thermal evolution described here.

\section{Conclusions and Discussions}

 We performed numerical simulations of the early phase of the low-mass star formation process just before the second collapse starts. Our simulation used a newly developed three-dimensional nested-grid self-gravitational FLD RMHD simulation code without assuming artificial thermal evolution. In our case, radiation transfer does not seem to change the qualitative scenario of low-mass star formation drastically. However, the temperature distribution is significantly changed by introducing radiation transfer. Realistic treatment of gas thermodynamics alters some properties and structure of the core quantitatively. The mass and size of the first core at a certain central density are larger because of a higher entropy, and the lifetime of the first core becomes slightly longer. Furthermore the envelope and the outer region of the first core become hotter. We suggest that the observational probability of such very young star forming sites rises when compared to previous predictions based on the simulations with the barotropic approximation.

 Barotropic approximation fails to reproduce realistic thermal evolution; needless to say radiation transfer plays a critical role there even in the low-mass regime. In the barotropic approximation all the gas elements trace the thermal evolution track of the gas at the center of the cloud which experiences no shock and, therefore, has minimum entropy in the spherically symmetric simulation. In this sense barotropic approximation tends to underestimate the gas entropy and temperature. In 3D, the RMHD simulation has more striking and complex differences from the barotropic simulation. On the other hand if there exist fast initial rotation and ineffective angular momentum transfer, then the entropy of the resulting thin disk-like first core may be lower than in the barotropic EOS. Thus radiation transfer will affect the mechanical properties such as stability or fragmentation through the thermal property of the gas. Determining a realistic temperature distribution is essential for predicting or interpreting the properties acquired in both the molecular emission lines and thermal continuum observations, for strong temperature dependences of the emissivity. Chemical reactions are also sensitive to temperature, so our results can be applied to a study of chemistry in molecular clouds and star forming regions.

 In our case, unlike previous MHD or RMHD simulations, there coexist two components of outflow from the first core. This feature suggests that the angular momentum transfer may occur necessarily not in single but in multiple stages, at least under a certain condition of magnetic fields although the situation depends on the initial condition and will be changed when magnetic dissipation is incorporated. Nonetheless our results imply a rich diversity of the outflow and first core. 

 Though on-going star formation like this simulation has never been observed directly, new telescopes are expected to reveal them, and ALMA seems to be the most promising in the near future. We propose some VeLLOs (Very Low Luminosity Objects; \citet{dnh08}) are candidates for such obscured star forming cores. Compact ($\sim 100 \ {\rm AU}$ scale) and warm ($\sim 30 \ {\rm K}$) molecular outflow can also be a good indicator of the first core. Our results can be compared with future observation.\\

 We thank Prof. Shu-ichiro Inutsuka, Dr. Kazuyuki Omukai and Dr. Takashi Hosokawa for fruitful discussions. Numerical computations were performed on NEC SX-9 at Center for Computational Astrophysics, CfCA, of National Astronomical Observatory of Japan. This work is supported in part by the Ministry of Education, Culture, Sports, Science and Technology (MEXT), Grant-in-Aid for Scientific Research, 21244021, 16204012 (Tomisaka), 20540238 (TM), 20740115 (KO) and 21740136 (MM). K. Tomida is supported by the Research Fellowship from the Japan Society for the Promotion of Science (JSPS) for Young Scientists.

\newpage


\end{document}